\pdfoutput=1
\documentclass[runningheads]{llncs}
\usepackage[T1]{fontenc}
\usepackage{newtxtext}
\usepackage[varvw]{newtxmath}
\usepackage{graphicx}
\usepackage{array}
\usepackage{booktabs}
\usepackage{marvosym}
\usepackage[nospace]{cite}
\usepackage[protrusion=false]{microtype}
\usepackage[hidelinks]{hyperref}
\hypersetup{pdftitle={MT-STAR: Spatio-Temporally Aware Generative Recommendation}}
\begin{document}
\title{MT-STAR: Spatio-Temporally Aware Generative Recommendation for Meituan's Local-Life Services}
\titlerunning{Spatio-Temporally Aware Generative Recommendation}
\author{Yingbin Mo\textsuperscript{\Letter} \and Wenjia Yi \and Qihang Li \and Peng Zhang \and Xiuqi Zheng \and Jialu Wang \and\\ Tao Feng \and Qingsong Hua \and Wei Lin}
\authorrunning{Y. Mo et al.}
\institute{Meituan, Beijing, China\\
\email{\{moyingbin,\allowbreak yiwenjia,\allowbreak liqihang05,\allowbreak zhangpeng115,\allowbreak zhengxiuqi,\allowbreak wangjialu10,\allowbreak fengtao02,\allowbreak huaqingsong,\allowbreak linwei31\}@meituan.com}}
\maketitle
\begin{abstract}

One practical application of generative recommendation (GR) is in local lifestyle services, where user needs are closely tied to location and time. In this context, semantic identifiers (SIDs) must capture both behavioral and geographic relationships, while the generator must account for preferences and accessibility that vary by location and time. Therefore, we present MT-STAR, a spatio-temporally aware generative recommendation framework with three components. First, we construct a collaborative geospatial SID (CoGeo-SID), whose prefixes reflect both geographic proximity and relationships inferred from user interactions. Second, we design a progressive learning method that aligns new contextual tokens with domain-specific knowledge and enhances the generator's ability to perceive user interests. Third, we introduce beam-guided preference optimization (BGPO), which adjusts the relative ranking of candidates based on behavioral preferences while taking into account request-dependent spatiotemporal accessibility. Offline experiments show a 35.79\% relative improvement in Order-Recall@30 over conventional dense retrieval. Incorporating the geographic penalty reduces the proportion of out-of-range candidates in the top 30 from 6.43\% to 2.57\%. In an online A/B test, MT-STAR achieves a 0.26\% relative uplift in UV\_CTCVR over the control group. MT-STAR has been fully deployed across Meituan's hotel business and now serves tens of millions of users.


\end{abstract}

\section{Introduction}

Generative recommendation generates item identifiers from user context and supports industrial recommendation \cite{onerec_v1,pinger,gr4ad} and search \cite{genstore,onesearch,oneretrieval}. Semantic identifiers (SIDs) encode items as short discrete sequences with shared prefixes \cite{tiger,forge}. Local-life services additionally require location and time to guide relevance and accessibility.

\begin{figure}[tb]
\centering
\includegraphics[trim=0 588 0 2,clip,width=0.72\linewidth]{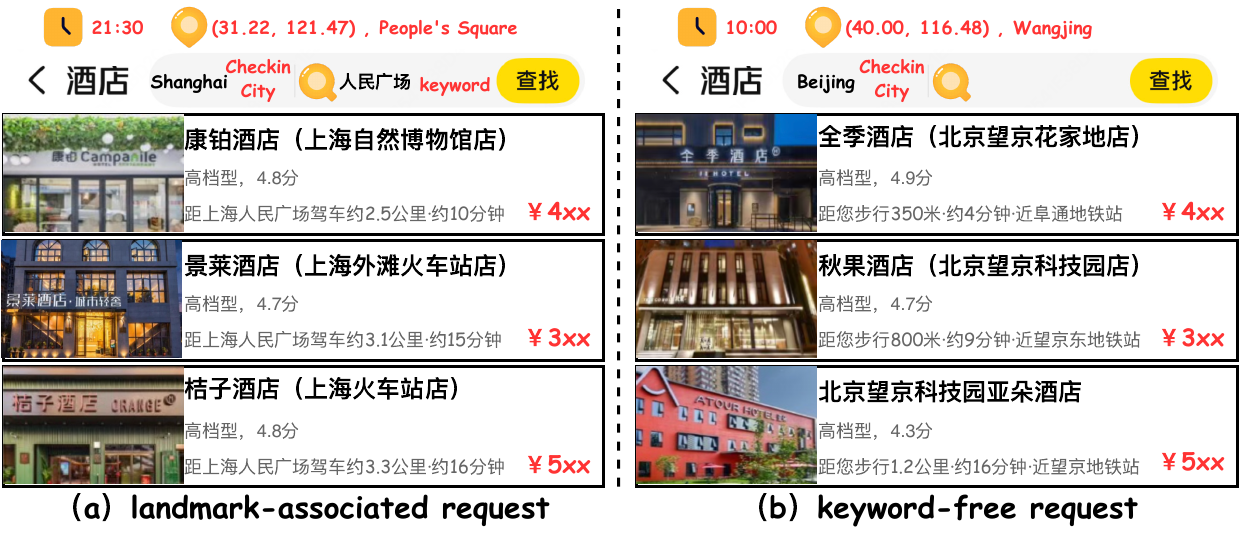}
\Description{Two Meituan list-page interfaces show a landmark-associated request and a keyword-free request, each with geographic and temporal context.}
\caption{Examples from Meituan's hotel search and recommendation engine}
\label{fig-application-scenarios}
\end{figure}

Meituan's hotel engine retrieves points of interest (POIs) relative to a request's geographic anchor: a destination landmark in landmark search or the user's location in keyword-free search (Fig.~\ref{fig-application-scenarios}). Preferences and time also affect relevance \cite{gnpr_sid,oneloc}.

Prior work incorporates geography into representations and codebooks \cite{gnpr_sid,geogr,progeo,lgsid} or generator training \cite{spar,oneloc}, regularizes codes with collaborative signals \cite{pinger,higr}, and aligns identifiers with language and behavior \cite{align3gr,lcrec,onerec_think}. Combining shared POI structure with request-dependent generation raises three challenges.

\textbf{(1) Retaining retrieval structure after discretization.} Combining text, behavior and geography in continuous embeddings does not ensure quantization preserves their relations. Yet early SID prefixes determine which candidate groups survive decoding.

\textbf{(2) Learning token meanings and contextual preferences.} New business tokens lack pretrained meanings, and item-level feedback leaves contextual preferences implicit, requiring supervision for token adaptation and request-level behavior.

\textbf{(3) Aligning candidate preferences with search and accessibility.} Finite beam search may discard a preferred POI before its complete SID is scored. Learning must guide both candidate and branch preferences, while geographic constraints must preserve genuine positive interactions.

We present \textbf{MT-STAR}, a spatio-temporally aware generative recommendation framework for Meituan's local-life services, with three contributions:

\begin{itemize}
\item \textbf{Collaborative-geospatial SID construction.} We combine collaborative text embeddings with multi-scale geo embeddings and regularize coarse quantization codes to preserve behavioral relevance and geographic proximity.
\item \textbf{Progressive contextual preference learning.} We align new business tokens with pretrained knowledge, then jointly train SID generation and auxiliary attribute prediction to learn request-dependent preferences.
\item \textbf{Beam-guided preference optimization.} We add observed positives missed by beam search, compare complete paths and their first divergent tokens, and penalize candidates outside the expected geographic range while preserving observed positives.
\end{itemize}

MT-STAR has been fully deployed across Meituan's hotel business, serving tens of millions of users. Offline experiments show a 35.79\% relative improvement in Order-Recall@30 over traditional dense retrieval. In the online A/B test, UV\_CTCVR increased by 0.26\% relative to the control.

\section{Problem Formulation}

\subsubsection{Task Definition and Autoregressive Generation.}

In SID-based generative POI recommendation, a residual-quantization encoder maps each POI's continuous embedding $\mathbf z(v)$ to a fixed-length SID $y=(y_1,\ldots,y_L)$ \cite{tiger}. Each request is represented as $x=(U,C_{\mathrm{hist}},A,C_{\mathrm{cur}})$, encompassing the user profile, historical context, behavior sequence, and current context. Given $x$, the policy $\pi_\theta$ retrieves POI candidates by generating high-probability SIDs in an autoregressive manner, $\pi_\theta(y\mid x)=\prod_{t=1}^{L}\pi_\theta(y_t\mid x,y_{<t})$. During supervised training, standard next-token prediction (NTP) minimizes $\mathcal L_{\mathrm{NTP}}(x,y^+)=-\sum_{t=1}^{L}\log\pi_\theta(y_t^+\mid x,y_{<t}^+)$ for an observed target SID $y^+$.

\subsubsection{Beam Decoding and Policy Optimization.}

At inference, beam search expands SID prefixes token by token, retaining the top-$B$ prefixes by cumulative log probability to decode at most $K$ POI candidates, where each completed path is scored by $s_\theta(y)=\log\pi_\theta(y\mid x)$. However, token-level NTP alone does not directly optimize candidate-level preferences or business constraints under finite-beam decoding. To bridge this gap, recent generative retrieval methods adopt PPO-style reinforcement learning \cite{ppo,onerec_v1,rere,unipolicy}, where beam search also supplies competitive training candidates \cite{localsug} and updates are governed by the token probability ratio $r_t(\theta)=\pi_\theta(y_t\mid x,y_{<t})/\pi_{\theta_{\mathrm{old}}}(y_t\mid x,y_{<t})$ with clipping range $\epsilon$. Building on this paradigm, BGPO introduces request-aware candidate augmentation and multi-granular credit assignment.

\section{Method}

\subsection{Overview of MT-STAR}

\begin{figure}[tb]
\centering
\includegraphics[trim=35 511 0 5,clip,width=\linewidth]{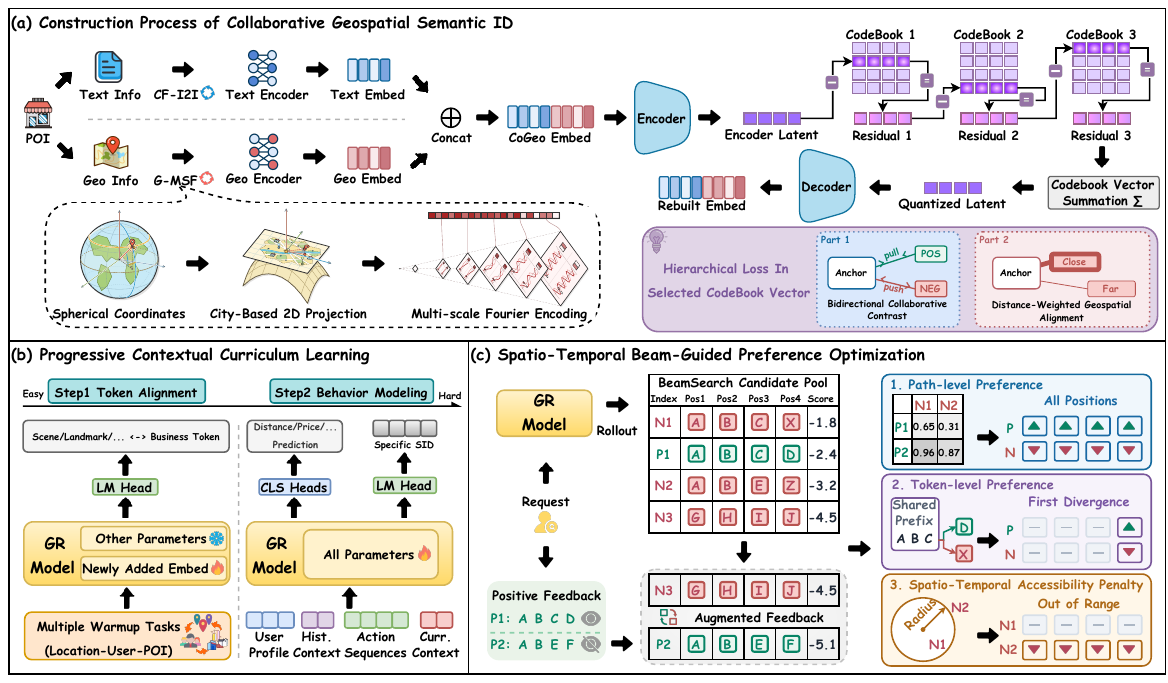}
\Description{MT-STAR has three components. Panel (a) combines text and geographic embeddings and preserves collaborative and geographic relationships during quantization. Panel (b) uses token alignment followed by behavioral modeling. Panel (c) optimizes beam-generated candidates using path-level preferences, token-level preferences, and request-dependent geographic accessibility.}
\caption{The overall framework of MT-STAR}
\label{fig-overview}
\end{figure}

Figure~\ref{fig-overview} illustrates the overall framework of MT-STAR, which is broadly divided into three steps. Step (a) merges text and geographic vectors and preserves behavioral and spatial relationships during coarse-grained encoding to construct a collaborative-geospatial SID. Step (b) introduces progressive context-based knowledge learning to establish associations between new business tokens and pretrained knowledge. Step (c) uses BGPO to refine candidate selection through path-level and token-level comparisons together with request-specific geographic constraints.

\subsection{Collaborative-Geospatial SID Construction}

\subsubsection{Collaborative-Geospatial Representation.}

We learn complementary text and geo embeddings to capture behavioral relevance and geography \cite{geogr,lgsid,spar,progeo}.

\paragraph{Collaborative Embedding.}

We fine-tune Qwen3-Embedding-8B on collaborative POI pairs with contrastive learning. Matryoshka representation learning \cite{mrl} optimizes nested embedding prefixes; a low-dimensional prefix serves as the text embedding $\mathbf e_v^{\mathrm{text}}$.

\paragraph{Geospatial Embedding.}

Our geographic multi-scale Fourier (G-MSF) encoder applies multi-scale Fourier features \cite{rff} to city-local coordinates, combines them with a city embedding into $\mathbf e_v^{\mathrm{geo}}$, and fits its similarity to geographic distance:
\begin{equation}
\begin{aligned}
\kappa(d)&=\frac1S\sum_{s=1}^{S}e^{-d^2/(2\ell_s^2)},\\
\mathcal L_{\mathrm{dist}}&=\mathbb E_{v,w}
\bigl[(\cos(\mathbf e_v^{\mathrm{geo}},\mathbf e_w^{\mathrm{geo}})-\kappa(d(v,w)))^2\bigr].
\end{aligned}
\end{equation}
Here, $d$ is Haversine distance, $\ell_s$ controls spatial scale, and the expectation is over sampled POI pairs. The decreasing target encourages higher similarity for nearby POIs than for distant POIs. The quantizer input is $\mathbf z(v)=\operatorname{Norm}([\mathbf e_v^{\mathrm{text}};\mathbf e_v^{\mathrm{geo}}])$, with L2 normalization applied after concatenation.

\subsubsection{Collaborative-Geospatial RQ-VAE.}

Reconstruction alone does not preserve relationships between POIs. We add two relation losses to RQ-VAE \cite{tiger,pinger}, constraining codeword vectors $\mathbf c_l(v)$ selected at coarse levels $l\in\mathcal C$ while leaving finer codes to retain POI-specific details.

\paragraph{Bidirectional Collaborative Contrast.}

For $N$ collaborative pairs $(a_i,p_i)$, define $S^{(l)}_{ij}=\cos(\mathbf c_l(a_i),\mathbf c_l(p_j))/\tau$. We align matched pairs in both directions \cite{higr}:
\begin{equation}
\mathcal L_{\mathrm{cf}}=\frac1{2|\mathcal C|}\sum_{l\in\mathcal C}\alpha_l
\left[\operatorname{CE}(S^{(l)},\mathbf t)+\operatorname{CE}((S^{(l)})^\top,\mathbf t)\right].
\end{equation}
Here, $\mathbf t=(1,\ldots,N)$ identifies matched pairs, CE averages row-wise cross-entropy, $\tau>0$ is a temperature, and $\alpha_l$ weights levels. Other batch entries serve as negatives.

\paragraph{Distance-Weighted Geospatial Alignment.}

For nearby POIs, $w_{vw}=e^{-d(v,w)/\ell_g}$ sets both a proximity target and a pair weight, with decay scale $\ell_g$:
\begin{equation}
\mathcal L_{\mathrm{geo}}^{(l)}(v,w)
=w_{vw}\left(\frac{1+\cos(\mathbf c_l(v),\mathbf c_l(w))}{2}-w_{vw}\right)^2.
\end{equation}
The rescaled cosine matches the target range $[0,1]$, and closer pairs have greater influence. We sum pair losses, normalize by total pair weight at each level, and average across coarse levels to obtain $\mathcal L_{\mathrm{geo}}$. The tokenizer objective is
\begin{equation}
\mathcal L_{\mathrm{tok}}=\mathcal L_{\mathrm{rec}}+\lambda_q\mathcal L_q
+\lambda_{\mathrm{cf}}\mathcal L_{\mathrm{cf}}+\lambda_{\mathrm{geo}}\mathcal L_{\mathrm{geo}},
\end{equation}
where $\mathcal L_q$ contains codebook and commitment losses.

\subsection{Progressive Contextual Preference Learning}

\subsubsection{Token Alignment.}

New business tokens include identifiers and context codes with no established meanings in the pretrained model. We first connect them to pretrained knowledge through domain-specific conditional generation \cite{lcrec,align3gr}. Training examples associate POI descriptions, location and request context with business tokens using aggregated behavioral relations. We train new token representations while freezing pretrained parameters \cite{onerec_think}, separating token adaptation from behavioral modeling.

\subsubsection{Behavioral Modeling.}

We then unfreeze the model and train request-conditioned SID generation. To make contextual preferences explicit, auxiliary heads predict attributes derived from the target POI and request, using $\mathbf h_x$, the request state before any target SID token. The objective is
\begin{equation}
\mathcal L_{\mathrm{SFT}}=\mathcal L_{\mathrm{SID}}+\eta(s)\mathcal L_{\mathrm{aux}},
\end{equation}
where $\mathcal L_{\mathrm{SID}}$ is weighted NTP, $\mathcal L_{\mathrm{aux}}$ combines weighted auxiliary cross-entropies, and $\eta(s)$ gradually introduces auxiliary supervision at training step $s$.

\subsection{Spatio-Temporal Beam-Guided Preference Optimization}

\subsubsection{Candidate Construction.}

We construct $\mathcal G_x$ by current-policy beam search \cite{localsug,rere,oneloc} and add missed clicked or ordered SIDs \cite{cq_sid}. For observed positives $\mathcal S_x^+$, let $\mathcal P_x=\mathcal G_x\cap\mathcal S_x^+$ and $\mathcal N_x=\mathcal G_x\setminus\mathcal S_x^+$. The actor teacher-forces all paths; its log probabilities, rather than search scores, determine preference credits and PPO ratios.

\subsubsection{Path-Level and Token-Level Preferences.}

We compare complete paths \cite{oneloc,rere} and individual branch choices \cite{hrpo,pinger,onesearch_v2}. For a candidate SID $y$, its complete-path score is $s_\theta(y)$, which is used for path-level comparison. For a token $a$ following prefix $u$, let $q_\theta(a\mid x,u)=\log\pi_\theta(a\mid x,u)$ denote its token-level log-probability. We build a prefix tree from all observed positive SIDs for the request. For each non-positive SID, we find the first token $w$ that leaves this tree and compare it with a token $c$ that continues a positive path from the same prefix $u$. Let $\mathcal Q_x$ contain distinct comparisons $(u,c,w)$. With path and token margins $m_{\mathrm{path}},m_{\mathrm{token}}\ge 0$, smoothing scales $\tau_{\mathrm{path}},\tau_{\mathrm{token}}>0$, and $\operatorname{sp}(z)=\log(1+e^z)$, the losses are
\begin{equation}
\begin{aligned}
\mathcal L_{\mathrm{path}}&=\frac{\tau_{\mathrm{path}}}{|\mathcal P_x||\mathcal N_x|}
\sum_{p\in\mathcal P_x,n\in\mathcal N_x}
\operatorname{sp}\!\left(\frac{s_\theta(n)-s_\theta(p)+m_{\mathrm{path}}}{\tau_{\mathrm{path}}}\right),\\
\mathcal L_{\mathrm{token}}&=\frac{\tau_{\mathrm{token}}}{|\mathcal Q_x|}
\sum_{(u,c,w)\in\mathcal Q_x}
\operatorname{sp}\!\left(\frac{q_\theta(w\mid x,u)-q_\theta(c\mid x,u)+m_{\mathrm{token}}}{\tau_{\mathrm{token}}}\right).
\end{aligned}
\end{equation}
If a comparison set is empty, its corresponding loss and credit are defined as zero. The pairwise losses above are used only to derive preference credits. We compute their gradients with the old policy frozen and use the negative gradients as fixed token-level credits. Path-level credits are broadcast to all tokens of the SID, whereas token-level credits are applied only at the first divergent position. This separates whole-path preference from branch-level preference without penalizing valid positive branches.

\subsubsection{Request-Dependent Geographic Accessibility.}

Generated candidates should remain geographically compatible with the request. During training, we estimate a geographic range $R_x$ for each request from the distances between the request anchor and its observed positive POIs, together with historical interaction distances for similar locations, times, and request types. Let
\begin{equation}
\mathcal N_x^{\mathrm{out}}=\{n\in\mathcal N_x:d_x(n)>R_x\},
\end{equation}
where $d_x(n)$ is the distance between candidate $n$ and the request anchor. Using the complete-path score $s_\theta$ defined above, we introduce the Geographic Accessibility loss:
\begin{equation}
\mathcal L_{\mathrm{GA}}(x)
=\frac{1}{C_x}\sum_{y\in\mathcal G_x}
m_x(y)\left[-\log\left(1-P_\theta(y\mid x)\right)\right],
\end{equation}
where $P_\theta(y\mid x)=\exp(s_\theta(y))$, $m_x(y)=\mathbf 1[y\in\mathcal N_x^{\mathrm{out}}]$, and $C_x$ is the number of candidates with valid geographic supervision. We set this term to zero when $C_x=0$. This sequence-level unlikelihood term reduces the complete-path probability of out-of-range non-positive candidates, while observed positive paths are exempt.

\subsubsection{Joint Objective.}

For each candidate in $\mathcal G_x$, let $A_t=A_t^{\mathrm{path}}+A_t^{\mathrm{token}}$ denote the combined fixed credit at token position $t$. We minimize the negative clipped PPO objective over $\mathcal G_x$:
\begin{equation}
\mathcal L_{\mathrm{clip}}(\theta)
=-\mathbb E_t\left[
\min\left(
r_t(\theta)A_t,
\operatorname{clip}(r_t(\theta),1-\epsilon,1+\epsilon)A_t
\right)
\right].
\end{equation}
We combine it with a positive NLL anchor \cite{pinger} and the geographic loss:
\begin{equation}
\mathcal L_{\mathrm{BGPO}}=\mathcal L_{\mathrm{clip}}
+\lambda_n\mathcal L_{\mathrm{NLL}}+\lambda_{\mathrm{GA}}\mathcal L_{\mathrm{GA}}.
\end{equation}
Here, $\mathcal L_{\mathrm{NLL}}=-|\mathcal S_x^+|^{-1}\sum_{p\in\mathcal S_x^+}s_\theta(p)$. Requests without a valid positive path are skipped for the NLL term, while requests without a valid range or geographic supervision contribute zero to $\mathcal L_{\mathrm{GA}}$. All positives are teacher-forced; NLL and geographic losses update the actor directly.

\section{Offline Experiments}

\subsection{Experimental Setup}

\subsubsection{Dataset.}

The MT-STAR dataset is constructed based on the natural traffic requests of Meituan's hotel list page. The data span eight consecutive days, with the first seven days used for training and the eighth day for testing. After processing and scene-specific down-sampling, the training set contains about 15 million requests and the test set contains about 200 thousand requests. To further evaluate the different objectives, we split the test set into click and order subsets by behavior type.

\subsubsection{Baselines.}

We compare MT-STAR with DSSM \cite{dssm}, a classic embedding-based retrieval (EBR) method, and TIGER \cite{tiger}, a foundational generative retrieval (GR) method. We also include two generative models designed for POI recommendation. GNPR-SID \cite{gnpr_sid} encodes each POI's region and typical visit times in its identifier and predicts the next POI from timestamped visits. GeoGR \cite{geogr} learns POI representations from geographically constrained co-visits and uses model predictions to refine identifier assignments.

\subsubsection{Metrics.}

We report query-macro Recall@30/100 and NDCG@30 for Click and Order under the same evaluation protocol (beam width = 100). Recall measures behavior-specific label coverage, and NDCG measures the rank of retrieved labels.

\subsubsection{Implementation.}

Each SID contains four positions: a fixed city marker followed by three residual-quantization markers, with 1,024 codewords at each quantization layer. The generator is a decoder-only model with 0.5B parameters, pretrained on Meituan Local-Life Services data. It consists of 8 Transformer layers, with a hidden dimension of 1,536, 24 attention heads, and 12 key-value heads. Progressive Contextual Curriculum Learning runs on 16 $\times$ NVIDIA A100 GPUs with a global batch size of 4,608. BGPO runs on 32 A100 GPUs, assigning 12 requests to each GPU per update and using no KL regularization. For final evaluation, we use the same beam search decoding protocol as the online system, with a beam width of 100.

\subsection{Experimental Results}

\subsubsection{Main Results.}

MT-STAR demonstrates the best recall and NDCG on all six metrics (Table~\ref{tab-offline-main}). All four GR methods outperform DSSM on this benchmark. GNPR-SID and GeoGR also improve over TIGER on all six metrics, showing the benefit of incorporating spatial or temporal signals into generative retrieval.

\begin{table}[tb]
\centering
\scriptsize
\setlength{\tabcolsep}{2.2pt}
\renewcommand{\arraystretch}{0.86}
\caption{Offline comparison of MT-STAR and baselines. Bold and underlined values indicate the best and second-best results in each column, respectively.}
\label{tab-offline-main}
\begin{tabular}{@{}cccccccc@{}}
\toprule
Paradigm & Method & \shortstack{Click\\Recall@30} & \shortstack{Click\\Recall@100} & \shortstack{Click\\NDCG@30} & \shortstack{Order\\Recall@30} & \shortstack{Order\\Recall@100} & \shortstack{Order\\NDCG@30}\\
\midrule
EBR & DSSM & 61.31\% & 74.12\% & 35.32\% & 64.96\% & 69.21\% & 44.35\%\\
GR & TIGER & 73.87\% & 86.36\% & 47.07\% & 85.93\% & 92.13\% & 55.92\%\\
GR & GNPR-SID & 74.85\% & \underline{87.52\%} & 48.33\% & 86.42\% & \underline{93.58\%} & 56.32\%\\
GR & GeoGR & \underline{75.01\%} & 87.49\% & \underline{48.57\%} & \underline{86.67\%} & 93.42\% & \underline{56.39\%}\\
GR & MT-STAR & \textbf{76.42\%} & \textbf{89.02\%} & \textbf{49.90\%} & \textbf{88.21\%} & \textbf{95.82\%} & \textbf{57.83\%}\\
\bottomrule
\end{tabular}
\end{table}

\subsubsection{Ablation Study.}

We start from Vanilla GR and introduce CoGeo-SID, the progressive curriculum, and BGPO in sequence (Table~\ref{tab-offline-ablation}). CoGeo-SID accounts for the largest gains in both Click and Order Recall@30. The progressive curriculum further improves recall for both behaviors using the same identifiers, suggesting that candidate coverage benefits from both how POIs are encoded and how the generator learns to use those codes. BGPO's clearest gain is in Order NDCG@30, with little change in recall. This pattern suggests that preference optimization places relevant POIs higher in the candidate list.

\begin{table}[tb]
\centering
\scriptsize
\setlength{\tabcolsep}{1.2pt}
\renewcommand{\arraystretch}{1.12}
\caption{Cumulative contributions of the three MT-STAR modules}
\label{tab-offline-ablation}
\begin{tabular*}{\linewidth}{@{\extracolsep{\fill}}>{\centering\arraybackslash}m{2.38cm}c>{\centering\arraybackslash}m{4.12cm}ccc@{}}
\toprule
Model & SID & Generator & Order & Order & Click\\[-1.5pt]
Configuration & Construction & Training & Recall@30 & NDCG@30 & Recall@30\\
\midrule
Vanilla GR & Text-only RQ-VAE & Standard SID prediction & 85.93\% & 55.92\% & 73.87\%\\
CoGeo-SID (a) & CoGeo-SID & Standard SID prediction & 87.35\% & 56.36\% & 75.79\%\\
Curriculum (a + b) & CoGeo-SID & Token alignment $\rightarrow$ behavioral modeling & 88.15\% & 56.67\% & \textbf{76.44\%}\\[2pt]
BGPO \mbox{(a + b + c = MT-STAR)} & CoGeo-SID & Token alignment $\rightarrow$ behavioral modeling \mbox{$\rightarrow$ BGPO} & \textbf{88.21\%} & \textbf{57.83\%} & 76.42\%\\
\bottomrule
\end{tabular*}
\end{table}

\paragraph{(a) Spatial and Collaborative Structure of SID Prefixes.}

We group POIs by all but the final SID token, keeping input embeddings and codebook sizes fixed. Cluster spatial dispersion (CSD) is the mean root-mean-square radius of non-singleton clusters, using geohash6 cell centers. For collaborative neighbor retention (CNR@5), we measure what fraction of each POI's valid top-five neighbors from interaction logs share its prefix, then average over POIs with valid neighbors.

CoGeo-SID balances spatial compactness and collaborative neighbor retention (Table~\ref{tab-sid-ablation}). Without bidirectional collaborative contrast (BCC), the mean cluster radius shrinks by 0.38 km, but retention drops by 5.55 percentage points. Without distance-weighted geospatial alignment (DWGA), retention rises by just 0.87 points while the mean radius more than doubles.

\begin{table}[!htb]
\centering
\scriptsize
\begin{minipage}[t]{0.42\linewidth}
\centering
\setlength{\tabcolsep}{4pt}
\renewcommand{\arraystretch}{1.05}
\caption{Spatial compactness and collaborative neighbor retention for SID prefixes}
\label{tab-sid-ablation}
\begin{tabular}{@{}ccc@{}}
\toprule
Quantizer & CSD (km) $\downarrow$ & CNR@5 (\%) $\uparrow$\\
\midrule
CoGeo-SID & \underline{2.44} & \underline{19.26}\\
w/o BCC & \textbf{2.06} & 13.71\\
w/o DWGA & 5.29 & \textbf{20.13}\\
\bottomrule
\end{tabular}
\end{minipage}\hfill
\begin{minipage}[t]{0.55\linewidth}
\centering
\setlength{\tabcolsep}{2.8pt}
\renewcommand{\arraystretch}{1.05}
\caption{Mean distance from each target POI to its nearest top-30 candidate}
\label{tab-location-ablation}
\begin{tabular}{@{}lcc@{}}
\toprule
Training variant & Original (km) $\downarrow$ & Replacement (km) $\downarrow$\\
\midrule
With token alignment & \textbf{2.16} & \textbf{5.93}\\
Without token alignment & 4.36 & 6.22\\
\bottomrule
\end{tabular}
\end{minipage}
\end{table}

\paragraph{(b) Geographic Anchors in POI Retrieval.}

For each request, we evaluate models trained with and without token alignment using the original location tokens and with the same replacement tokens sampled from another grid cell in the same city. For each target POI, we measure its distance to the nearest of the model's 30 highest-ranked candidates. We average these distances first within each request and then across requests.

With the original location tokens, alignment reduces the mean distance to the nearest candidate by 2.20 km. This advantage narrows to 0.29 km after same-city replacement, suggesting that alignment helps the retriever use the request's location to find candidates closer to the target POI (Table~\ref{tab-location-ablation}).

\paragraph{(c) Geographic Accessibility in BGPO.}

We compare BGPO with and without its geographic accessibility objective (GA). For each request, we measure the fraction of returned POIs that fall outside the request's geographic range, using up to 30 candidates. FarRatio@30 is the average of this fraction across requests. For the post-hoc baseline, we filter candidates generated without GA and tune the distance threshold on the validation set to match the FarRatio@30 of BGPO with GA as closely as possible.

\begin{table}[tb]
\centering
\scriptsize
\setlength{\tabcolsep}{4pt}
\renewcommand{\arraystretch}{0.96}
\caption{Geographic accessibility and order recall under different candidate-selection strategies}
\label{tab-far-ratio}
\begin{tabular}{@{}lcc@{}}
\toprule
Candidate selection & FarRatio@30 $\downarrow$ & Order Recall@30 $\uparrow$\\
\midrule
BGPO with GA & \underline{2.57\%} & \textbf{88.21\%}\\
BGPO w/o GA & 6.43\% & 88.07\%\\
BGPO w/o GA + distance filter & \textbf{2.55\%} & \underline{88.13\%}\\
\bottomrule
\end{tabular}
\end{table}

As shown in Table~\ref{tab-far-ratio}, BGPO with GA retained slightly higher order recall than the post-hoc filter at comparable out-of-range rates.

\begin{figure}[!htb]
\centering
\includegraphics[trim=0 455 0 0,clip,width=0.60\linewidth]{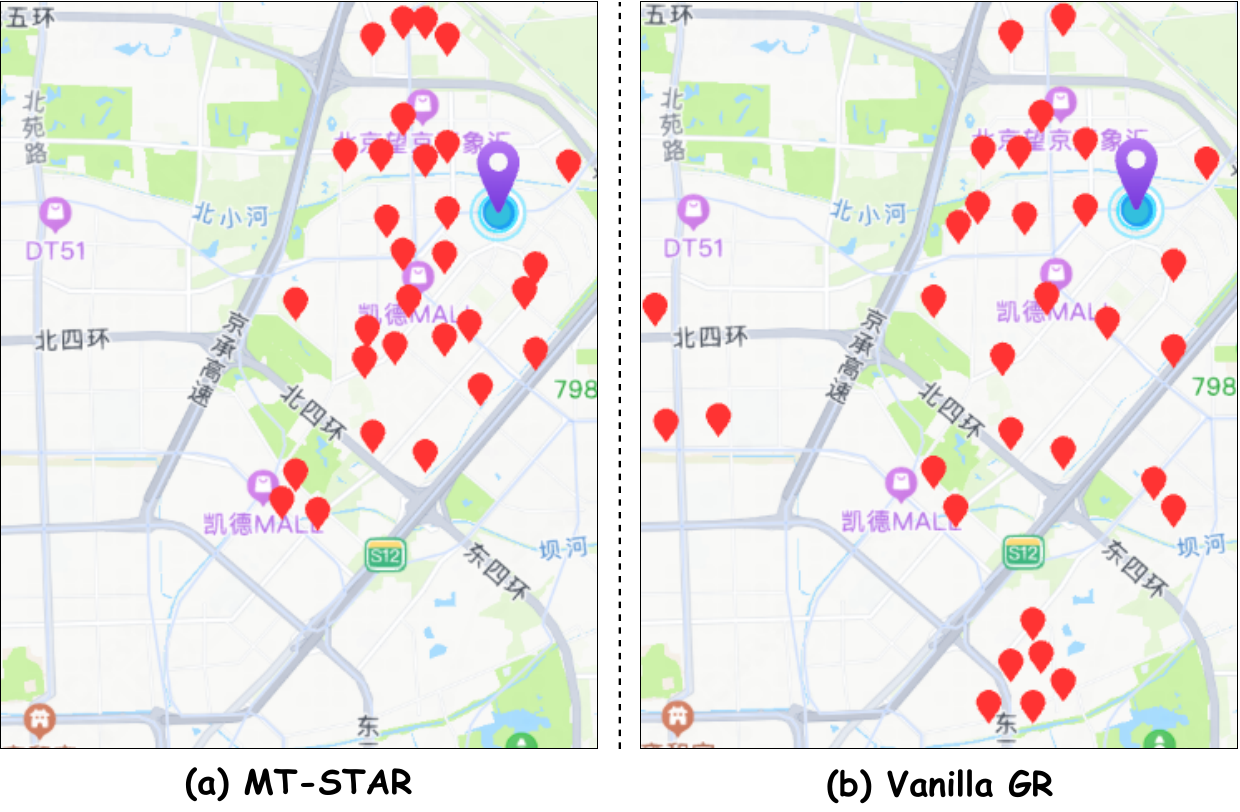}
\Description{Maps of candidates for the same request show MT-STAR on the left and Vanilla GR on the right. MT-STAR candidates are more concentrated around the geographic anchor, while Vanilla GR includes more distant outliers.}
\caption{Comparison of recall results for the same request}
\label{fig-case-study}
\end{figure}

\subsubsection{Case Study.}

Figure~\ref{fig-case-study} shows that MT-STAR returns a more compact cluster around the anchored location, while Vanilla GR also retrieves several isolated POIs farther away.

\section{Online Performance}

\subsection{Online Deployment}

\subsubsection{Inference Engine.}

MT-STAR serves Meituan's hotel search on the SLM inference engine with beam width 100 and FP8 mixed precision. Short, fixed-length SIDs enable a fixed execution graph: prompt prefill, beam candidate selection, and state updates are compiled into one generation node run by a C++ runtime. Within each request, beams share prompt key--value states and retain separate suffix states, reducing replication and data movement. Additionally, preallocated tokenization and encoding buffers reduce non-model overhead. Table~\ref{tab-inference-latency} shows that the SLM engine reaches 41.71\% MFU and achieves lower average and P90 latency at both tested loads.

\begin{table}[tb]
\centering
\scriptsize
\begin{minipage}[t]{0.52\linewidth}
\centering
\setlength{\tabcolsep}{3pt}
\renewcommand{\arraystretch}{1.08}
\caption{Inference latency (ms) and effective model FLOPs utilization (MFU, \%) on L40}
\label{tab-inference-latency}
\begin{tabular}{@{}lccccc@{}}
\toprule
Engine & \shortstack{MFU\\(\%) $\uparrow$} & \shortstack{10 QPS\\Avg.} & \shortstack{10 QPS\\P90} & \shortstack{60 QPS\\Avg.} & \shortstack{60 QPS\\P90}\\
\midrule
SGLang & 6.82\% & 45 & 60 & 100 & 165\\
SLM & \textbf{41.71\%} & \textbf{11} & \textbf{14} & \textbf{19} & \textbf{33}\\
\bottomrule
\end{tabular}
\end{minipage}\hfill
\begin{minipage}[t]{0.45\linewidth}
\centering
\setlength{\tabcolsep}{6pt}
\renewcommand{\arraystretch}{1.08}
\caption{Relative changes in online metrics. An asterisk denotes $p<0.05$.}
\label{tab-online-ab}
\begin{tabular}{@{}cc@{}}
\toprule
Metric & Relative change\\
\midrule
OC & +1.74\%\textsuperscript{*}\\
UV\_CTCVR & +0.26\%\textsuperscript{*}\\
\bottomrule
\end{tabular}
\end{minipage}
\end{table}

\subsubsection{System Architecture.}

\begin{figure}[tb]
\centering
\begin{minipage}[t]{0.42\linewidth}
\centering
\includegraphics[trim=27 350 104 0,clip,height=4.1cm]{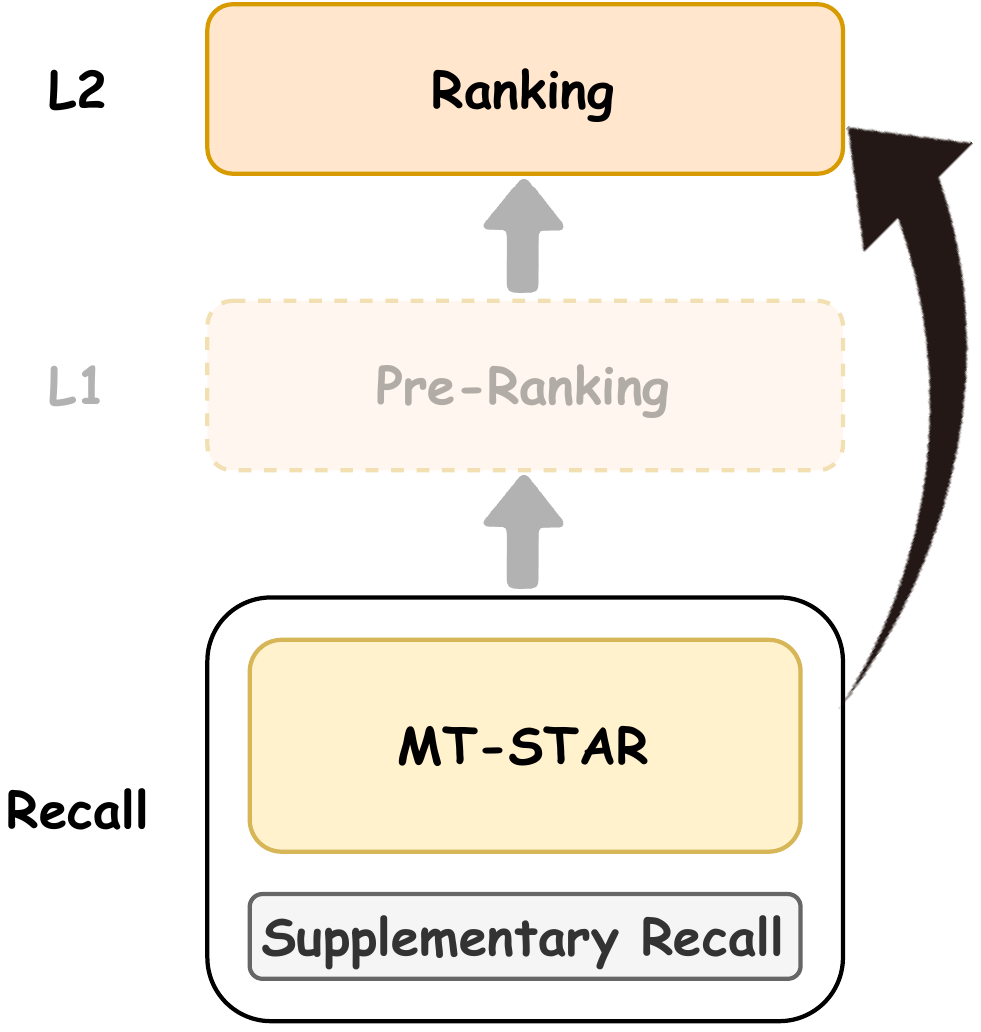}
\Description{The online architecture sends MT-STAR and supplementary recall candidates directly to final ranking while bypassing pre-ranking.}
\caption{System architecture of online MT-STAR serving}
\label{fig-online-architecture}
\end{minipage}\hfill
\begin{minipage}[t]{0.54\linewidth}
\centering
\includegraphics[trim=4 424 4 11,clip,height=4.1cm]{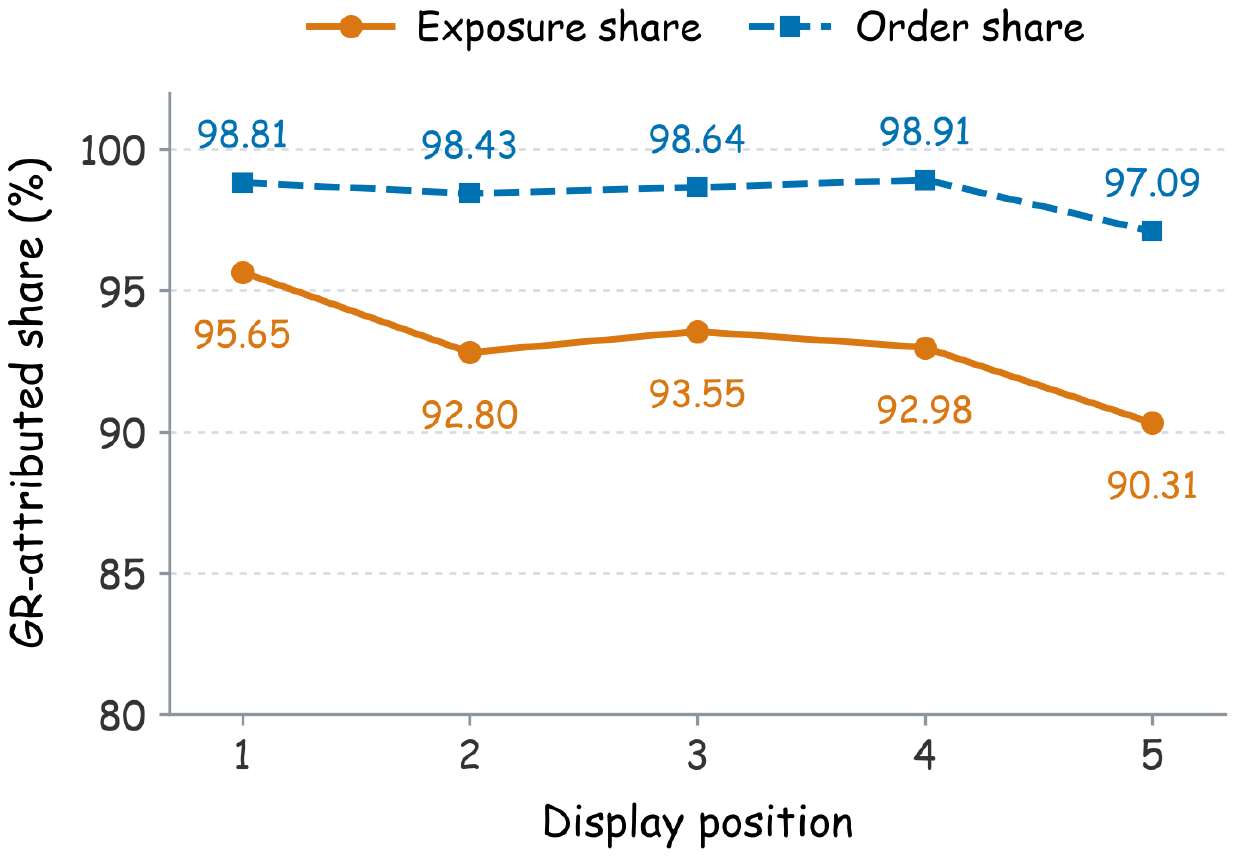}
\Description{MT-STAR exposure and order shares at the first five display positions exceed 90 percent and 97 percent, respectively.}
\caption{Exposure and order shares attributed to MT-STAR at the first five display positions}
\label{fig-gr-share-top5}
\end{minipage}
\end{figure}

MT-STAR serves as the primary recall channel and maintains order coverage, while SLM provides the low-latency inference engine for online serving. Figure~\ref{fig-online-architecture} shows the resulting architecture, which retains a small supplementary recall channel and sends the combined candidates directly to final ranking, bypassing pre-ranking. This removes intermediate candidate pruning and reduces retrieval and scoring overhead, allowing the latency budget to support a wider beam search while keeping end-to-end latency nearly unchanged.

\subsection{A/B Test}

\subsubsection{Online Metrics and Performance.}

We report two online metrics: order count (OC), the number of completed orders, and unique visitor conversion rate (UV\_CTCVR), the proportion of unique visitors who complete a transaction. UV\_CTCVR is the primary business metric. As shown in Table~\ref{tab-online-ab}, MT-STAR improves both metrics over the control.

Across the first five positions on the list page, candidates retrieved by MT-STAR account for more than 90\% of exposures and over 97\% of orders (Fig.~\ref{fig-gr-share-top5}). This concentration makes MT-STAR the primary recall source for these high visibility positions.

\section{Related Work}

Generative recommendation (GR) represents items with semantic identifiers and retrieves them by autoregressive decoding. TIGER introduced residual-quantized SIDs, and industrial systems such as OneRec and GenStore established this paradigm in recommendation and search \cite{tiger,onerec_v1,genstore}. Later work strengthens identifier construction and token adaptation through collaborative semantics, auxiliary objectives, codebook regularization, and staged training \cite{lcrec,align3gr,higr,onerec_think}. Local-service methods incorporate geographic or temporal signals into identifiers, representations, generation, or rewards \cite{gnpr_sid,geogr,progeo,spar,lgsid,oneloc}. Preference-oriented methods align training with beam decoding through candidate comparison, rank-sensitive rewards, and hierarchical credit assignment \cite{localsug,rere,hrpo,onesearch_v2}. These directions address complementary parts of local-service retrieval. MT-STAR treats them as a coupled retrieval problem, connecting collaborative-geospatial SID construction and business-token adaptation with request-dependent preference and accessibility control.

\section{Conclusion}

MT-STAR establishes a request-aware generative retrieval framework for local-life services. CoGeo-SID preserves collaborative and geographic relations in coarse codes, progressive contextual learning adapts new business tokens, and BGPO aligns candidate preferences with request-dependent accessibility. Offline and online evaluations show improvements in retrieval quality and business metrics. Future work will extend BGPO with process reward models for finer-grained supervision and explore unified objectives for multiple business goals, including relevance, accessibility, and coverage.

\begin{credits}

\subsubsection{\discintname} The authors have no competing interests to declare.

\end{credits}
\bibliographystyle{splncs04}
\bibliography{references}
\end{document}